\documentclass[10pt,aps,prb,twocolumn,showpacs,amssymb,floatfix]{revtex4-1}
\usepackage{amsmath,graphics,epsfig,mathrsfs}
\usepackage{bm}
\usepackage{epigraph}

\usepackage{color}

\usepackage{color}
\definecolor{darkgreen}{rgb}{0.0, 0.5, 0.0}
\definecolor{brown}{rgb}{0.59, 0.29, 0.0}
\definecolor{darkgreen2}{rgb}{0.01, 0.75, 0.24}
\usepackage{hyperref}
\hypersetup{backref=true,pagebackref=true,hyperindex=true,colorlinks=true,citecolor=blue, breaklinks=true,urlcolor=
blue,linkcolor=blue,bookmarks=true,bookmarksopen=true}

\begin{document}
\title{Topological Hall and Spin Hall Effects in Disordered Skyrmionic Textures}
\author{Papa Birame Ndiaye}
\author{Collins Ashu Akosa}
\author{Aur{\' e}lien Manchon$^{1}$}
\affiliation{$^1$King Abdullah University of Science and Technology (KAUST), Physical Science and Engineering Division, Thuwal 23955-6900, Saudi Arabia.}
\date{\today}

\begin{abstract}    
We carry out a throughout study of the topological Hall and topological spin Hall effects in disordered skyrmionic systems: the dimensionless (spin) Hall angles are evaluated across the energy band structure in the multiprobe Landauer-B\"uttiker formalism and their link to the effective magnetic field emerging from the real space topology of the spin texture is highlighted. We discuss these results for an optimal skyrmion size and for various sizes of the sample and found that the adiabatic approximation still holds for large skyrmions as well as for few atomic size-nanoskyrmions. Finally, we test the robustness of the topological signals against disorder strength and show that topological Hall effect is highly sensitive to momentum scattering.
\end{abstract}

\maketitle

\section{Introduction}
Since the discovery of the ordinary Hall effect \cite{Hall1879} (OHE) in 1879, closely related phenomena such as anomalous Hall effect \cite{Hall1881} (AHE) and spin Hall effect \cite{Dyakonov1971,Hirsch1999,Kato2004,Wunderlich2005,Valenzuela2006} (SHE) have been experimentally reported and their underlying mechanisms theoretically investigated \cite{Nagaosa2010,Sinova2015}. Their occurrence in a broad range of solids and electron gases under different conditions suggests a common denominator which is the conjunction of time-reversal symmetry breaking by either external magnetic field or magnetization and the onset of an effective Lorentz force either driven by external magnetic field or spin-orbit coupling (SOC). In ferromagnetic conductors for instance, where magnetization and SOC are present, AHE generates a transverse charge voltage at opposite edges of the sample \cite{Nagaosa2010}. In contrast, in normal metals or semiconductors where only SOC is present, SHE induces a chargeless spin voltage \cite{Kato2004,Wunderlich2005,Valenzuela2006}. In both AHE and SHE, SOC induces an effective Lorentz force, related either to a disorder-driven renormalization of the velocity operator or to the band structure Berry curvature \cite{Nagaosa2010,Sinova2015}. The anomalous velocity arises from the fictitious magnetic field ${\bf B}(\rm p)$ that emerges in momentum space. \par

Interestingly, this emergent magnetic field does not necessarily need to be in momentum space, but can also exist in real space \cite{Xiao2010,Yasuda2016}.
It is well known that when electrons flow in a non-trivial magnetic texture, they experience an emergent electromagnetic field \cite{Barnes2007,Saslow2007}. The emergent electric field $ E_{i}^s=(s \hbar/2e) {\bf m}\cdot(\partial_t{\bf m}\times\partial_i{\bf m})$ produces a spin motive force \cite{Yang2009,Tanabe2012}, i.e. a time-dependent magnetization ($\partial_t{\bf m}\neq0$) induces a local spin current \cite{Barnes2007,Saslow2007}. The emergent magnetic field ${\bf B}^s=(-s \hbar/2e){\bf m}\cdot(\partial_x{\bf m}\times\partial_y{\bf m}){\bf z}$ creates an effective Lorentz force\cite{Schulz2012} on the flowing electron that changes sign on the two opposite spins, creating a local, spin-dependent OHE. This emergent magnetic field, formed by the solid angle subtended by the magnetic moments of the spin texture \cite{Taguchi2001}, is capable of inducing the transverse motion of electrons like any real magnetic field giving rise to the so-called topological Hall effect \cite{Bruno2004} (THE) in magnetic textures with non-trivial topology. \par

The role of real space topology keeps on increasing since the experimental discovery of magnetic skyrmions \cite{Muhlbauer2009,Neubauer2009,Yu2010,Heinze2011, Jiang2015,Woo2016,Boulle2016,Chen2015,Moreau-Luchaire2016}, which are topologically non-trivial spin textures \cite{Nagaosa2013} in non-centrosymmetric ferromagnetic structures.
Skyrmions are in pole position in the racetrack memory search, thanks to prominent features that make them the ultimate bit of information \cite{Fert2013}: in contrast with magnetic domain walls, magnetic skyrmions are topological defects, localized in space, and present a decent robustness against pinning by magnetic defects, enabling current-driven motion at low current density. Different skyrmion sizes have already been reported in the bulk of B20 compounds or in magnetic multilayers with broken inversion symmetry. For instance, a skyrmion diameter of 70 nm has been obtained in thin film FeGe \cite{Yu2010}, as compared to 30 nm in ultrathin (Ir/Co/Pt)$_10$ multilayers \cite{Moreau-Luchaire2016} and 18 nm for MnSi \cite{Neubauer2009}, down to 1 nm in Fe monolayer deposited on Ir($111$) surface \cite{Heinze2011}. These small sizes correspond to emergent magnetic fields ranging from 1T to 4000T.

The topological properties of skyrmions ensure that the total flux generated by a single skyrmion equals one flux quantum, $h/e$. 
Recently, discretized topological Hall effect has been observed \cite{Kanazawa2015} in constricted geometry and the emergence of quantum AHE in a skyrmion crystal has been theoretically explored \cite{Hamamoto2015}. An intriguing topological spin Hall effect (TSHE) has been obtained numerically in a single skyrmion \cite{Yin2015}. This TSHE displays an atypical energy dependence that contrasts with the one of THE.\par

In this work, we focus on the topological electronic transport in ferromagnetic skyrmions, in both clean and disordered regimes. We use a tight-binding model to study charge-spin transport quantities in a ferromagnetic conductor perforated by an isolated skyrmion. In particular, we investigate the dimensionless charge and spin Hall angles quantifying the strength of THE and TSHE as a function of the carrier transport energy. We also test the magnitude of these two effects as a function of the skyrmion radius and find that the THE and TSHE reach their saturated values even for few-atom-size skyrmions. Finally, we inspect the robustness of THE and TSHE as a function of the disorder strength and find that the Hall effect is significantly reduced even when the mean free path is larger than the skyrmion radius. 

The paper is organized as follows: Section II presents the theoretical method and offers a general discussion about the charge and spin transport calculation in the tight-binding system. The numerical results for the clean and disordered regimes are presented and analyzed in Sec. III. 
Conclusion and perspectives are provided in Sec. IV.
\section{Model}
\subsection{Theoretical Method}
\begin{figure}[t]
\includegraphics[width=8.9cm]{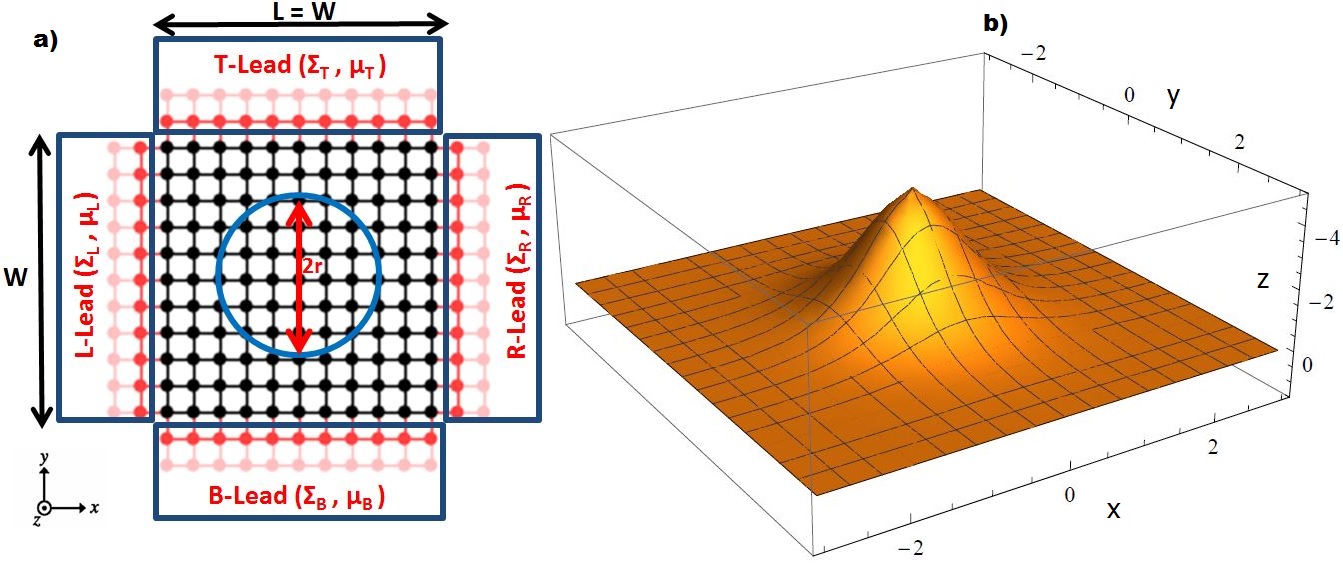}
\caption{\label{fig:1} (Color online) (a) Diagram of the 4-terminal setup, made up of a central skyrmion scattering region attached to four ferromagnetic leads L, T, R, B at chemical potentials $\mu_{L,T,R,B}$. 
A voltage bias is applied between L and R while the induced transverse charge-spin voltages are probed in T and B. (b) Magnetic field emerging from the skyrmion texture.} 
\end{figure} 
In this section, we present the system within a suitable framework to quantify the topological Hall effect and the topological spin Hall effect arising from the emergent magnetic field of the skyrmionic texture. 
This effective field, represented in Fig. \ref{fig:1}(b) reads $B_z({\bf r})=\frac{\hbar}{2e}{\bf m}\cdot(\partial_x{\bf m}\times\partial_y{\bf m})$ where the local magnetization unit vector is ${\bf m}=(\sin\theta(r)\cos\phi_{xy},\sin\theta(r)\sin\phi_{xy},\cos\theta(r))$, with $r=\sqrt{x^2+y^2}$ and $\phi_{xy}$ is the azimuthal angle. The total magnetic flux over the sample is simply $\int d^2{\bf r} B_z({\bf r}) =N \Phi_0$ where $N$ is the quantized topological winding number or Chern number and $\Phi_0$, the quantum of flux. 
We model our system, a thin ferromagnetic layer pierced by a single skyrmion at its center [see Fig. \ref{fig:1}(a)], as a two dimensional square lattice of size $L\times Wa_0^2$ with $a_0$ being the lattice constant, connected to four external semi-infinite ferromagnetic reservoirs  to ensure the continuity in magnetization between the central region of study and the leads. 
We adopt a skyrmionic profile given by the polar angle 
of  the local magnetization direction $\theta({\bf r})=4\tan^{-1} \exp(\frac{r}{r_0})$. We note that this differs from the \emph{linear} profiles discussed in Ref. \onlinecite{Yasuda2016} and from the Usov ansatz for skyrmions described in Ref. \onlinecite{Kim2015} for example. Nevertheless, the numerical differences are small and do not influence the qualitative features of the topological Hall angles as long as minimal distances between the sample edges and the skyrmion spatial extension are respected. 
Our considerations are based on steady state conditions for which  the skyrmion is pinned and its dynamics neglected as we consider very small injected currents.

We calculate the transport properties of interest using the wave function formulation of the scattering problem method as implemented in the software package Kwant \cite{Groth2014}. In a single-band tight-binding model, the physical quantities are expressed in the basis of the local atomic sites wave function and the electron Hamiltonian reads
\begin{equation}\label{Hamilt}
\hat{H}=\sum_i \hat{c}_i^\dagger\epsilon_i\hat{c}_i-t\sum_{\langle i,j\rangle}(\hat{c}_i^\dagger\hat{c}_{j}+{\rm H.c})
-\Delta\sum_i \hat{c}_i^\dagger{\bf m}_i\cdot \hat{\boldsymbol{\sigma}}\hat{c}_i
\end{equation}
where $\epsilon_i$ is the on-site energy, $t$ is the hopping parameter between the neighboring sites $i$ and $j$, $\Delta$ is the strength of the exchange 
coupling between the background magnetic texture ${\bf m}_i$ of the scattering region and the itinerant electron with spin represented by the vector of spin Pauli matrices $\hat{\boldsymbol{\sigma}}$. $\hat{c_i}^\dagger=(\hat{c}_{i\uparrow}^\dagger,\hat{c}_{i\downarrow}^\dagger)$ is the spinor form of the usual fermionic creation operator at the site $i$ ($\uparrow,\downarrow$ refer to the spin projection along the quantization axis, i.e $z$-axis). 
The local magnetic moment direction ${\bf m}_i$ on site $i=(i_x,i_y)$ of the scattering region is determined by the spatial extension of the skyrmion as ${\bf m}_i=\hat{z}$ everywhere except inside the skyrmion.
Neither external magnetic field nor SOC are considered in this work. This rules out not only OHE but also the 'conventional' AHE and SHE. Therefore, any Hall signal computed in this study arises solely from the topology of the magnetic texture.

For the coherent charge and spin transport calculation, we apply the Landauer-B\"uttiker formalism to the four-terminal cross bar device as shown in Fig. \ref{fig:1}(a), in which a voltage bias is added between the left lead (L) and the right lead (R), imposing a longitudinal flowing charge current. The induced transverse charge and spin currents are probed using the top lead (T) and the bottom lead (B).

\subsection{Landauer-B\"uttiker for charge and spin currents}
In our tight-binding model, we define each ferromagnetic lead in Fig. \ref{fig:1} as consisting of two leads allowing only one spin species $\uparrow,\downarrow$ to propagate. The tight-binding Hamiltonian in Eq. \eqref{Hamilt}, with the skyrmion texture,  mixes the two spin channels. Therefore, the implementation using Kwant provides directly the spin-resolved transmission coefficients within the standard multiprobe Landauer-B\"uttiker formalism \cite{Buttiker1986,Datta1995}. 
The electric currents $I_m^e$ in a structure attached to many leads (labeled by $m$= L, T, R, B) are calculated as
\begin{equation}\label{eq:current}
I_m^e = \frac{e^2}{2\pi\hbar}\sum_{n\ne m, \sigma, \sigma'}\left( T_{nm}^{\sigma'\sigma}V_m - T_{mn}^{\sigma\sigma'}V_n\right),
\end{equation}
where $T_{nm}^{\sigma'\sigma}$ is the transmission coefficient for an electron from lead-$m$ with spin $\sigma$ to lead-$n$ with spin $\sigma'$.
We note that the vector composed of the four terminal charge currents is straightforwardly written as a matrix of the transmission coefficients multiplied by the vector of the four lead voltages. The 4$\times$4 matrix associated with the linear system described by Eq. \eqref{eq:current} is obviously singular, because of the total charge current conservation at steady state ($T_{mm}^{\sigma'\sigma}= -\sum_{n\neq m} T_{mn}^{\sigma'\sigma}$). 
Therefore, we can without loss of generality set one of the voltage, $V_{\rm B} = 0$ and write
$\left(V_{\rm L}, V_{\rm T}, V_{\rm R}\right)^{\rm T} 
= \frac{2\pi\hbar}{e^2} \mathcal{R}\left( I_{\rm L}^e, I_{\rm T}^e, I_{\rm R}^e \right)^{\rm T}$
where $\mathcal{R}$ is the inverse of the 3$\times$3 transmission matrix, straightforwardly obtained from Eq. \eqref{eq:current}. When we enforce a small longitudinal charge bias between lead L and lead R i.e $\mu_{\rm R}-\mu_{\rm L}=e\delta V$, $I_{\rm L} = -I_{\rm R} = I$ and $I_{\rm T} = 0$ for the Hall measurements, the terminal voltages are expressed as
\begin{equation}
\begin{split}
 & V_{\rm L} =\left(\mathcal{R}_{\rm 11} - \mathcal{R}_{\rm 13}\right)\delta V/D, \\
& V_{\rm T} =\left(\mathcal{R}_{\rm 21} - \mathcal{R}_{\rm 23}\right)\delta V/D, \\
& V_{\rm R} =\left(\mathcal{R}_{\rm 31} - \mathcal{R}_{\rm 33}\right)\delta V/D, \nonumber
\end{split}
\end{equation}
and $V_{\rm B}=0$ with $D=\mathcal{R}_{\rm 11} + \mathcal{R}_{\rm 33} - \mathcal{R}_{\rm 13} - \mathcal{R}_{\rm 31}$ and $\delta V=\left(\mu_{\rm L} - \mu_{\rm R}\right)/e$ being the imposed voltage bias between the left and right leads.
The transverse Hall voltage and the topological Hall angle $\theta_{\rm TH}$ are readily evaluated as 
\begin{equation}
\theta_{\rm TH} = \frac{E_H}{E_x}=\frac{V_{\rm T}-V_{\rm B}}{V_{\rm R}-V_{\rm L}}.
\end{equation}
In order to calculate the spin Hall angle, we first define the quantities
\begin{eqnarray}
T_{ mn}^{\rm in} = T_{mn}^{\uparrow\uparrow} + T_{mn}^{\uparrow\downarrow} 
- T_{ mn}^{\downarrow\downarrow} - T_{mn}^{\downarrow\uparrow},\\
T_{mn}^{\rm out} = T_{mn}^{\uparrow\uparrow} + T_{mn}^{\downarrow\uparrow} 
- T_{mn}^{\downarrow\downarrow} - T_{mn}^{\uparrow\downarrow},
\end{eqnarray}
quantifying the spin current entering in and going out of the lead $m$. The different terminal spin currents are defined as \cite{Pareek2004,Nikolic2005}
\begin{equation}
I_m^s =\frac{e}{4\pi} \sum_{n\ne m}\left( T_{nm}^{\rm out}V_m - T_{mn}^{\rm in}V_n\right).
\end{equation}
For instance, the spin current in the left lead is $I_{\rm L}^s =\frac{e}{4\pi} \left( [T_{\rm TL}^{\rm out} + T_{\rm RL}^{\rm out} + T_{\rm BL}^{\rm out}]V_{\rm L} - T_{\rm LT}^{\rm in}V_{\rm T} - T_{\rm LR}^{\rm in}V_{\rm R} - T_{\rm LB}^{\rm in}V_{\rm B} \right)$.
From the spin and charge currents, we can calculate the topological spin Hall angle (TSH) as
\begin{equation}
\theta_{\rm TSH} = \frac{2e}{\hbar}\left(\frac{I_{\rm T}^s -I_{\rm B}^s}{I_{\rm L}^e -I_{\rm R}^e}\right).
\end{equation}
\section{Results and Analysis}
\subsection{Preliminary Results for a single skyrmion}
\begin{figure}[b]
\includegraphics[width=8.4cm]{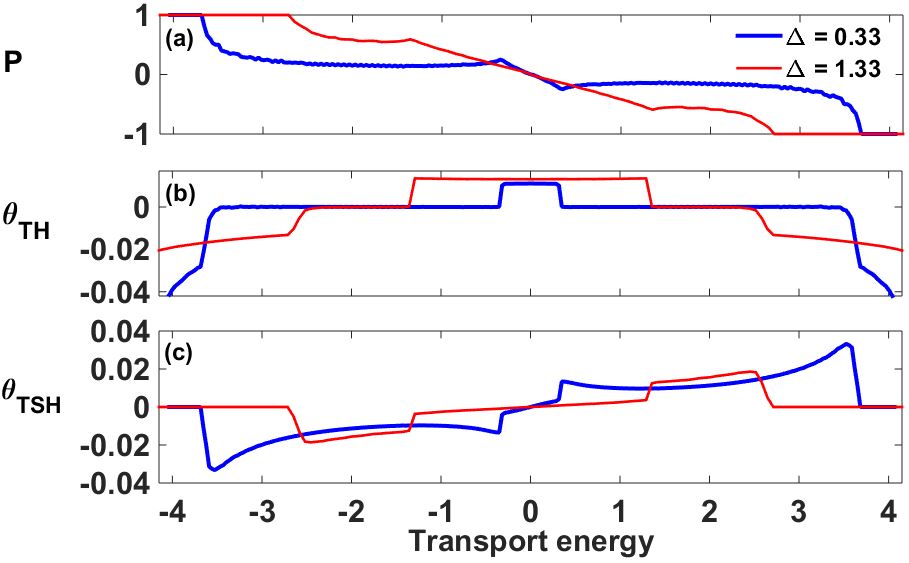}
\caption{\label{fig:2} (Color online) (a) The spin polarization and (b,c) topological Hall angles for different transport energies $\epsilon_{\rm tr}$ for a sample size of $L=W=96a_0$ and a skyrmion radius of 10$a_0$. The value of the exchange coupling $\Delta$ defines the boundaries and affects the magnitudes of $\theta_{\rm TH}$ and $\theta_{\rm TSH}$.}
\end{figure}

We first benchmark our model by computing $\theta_{\rm TH}$ and $\theta_{\rm TSH}$, quantifying the THE and TSHE respectively, as a function of different transport energy $\epsilon_{\rm tr}$, following Ref. \onlinecite{Yin2015}. The system we consider is a typical ferromagnetic metal, described in Eq. \eqref{Hamilt}, such that the splitting of the energy band structure $\Delta$ is much smaller than the tight-binding bandwidth $8t$: the spin-resolved energy bandwidths therefore overlap in some range of the transport energy $\epsilon_{\rm tr}$. All the energies are normalized to the hopping parameter $t$. 
The band structure extends from $\epsilon_{\rm tr}=-4t-\Delta$ to $\epsilon_{\rm tr}=4t+\Delta$, and the associated spin polarization P is displayed in Fig. \ref{fig:2}(a). Half metallic states (P=$\pm$1) are located at the edges of the band structure, specifically for $|\epsilon_{\rm tr}|>4t-\Delta$. In the range $-4t+\Delta<\epsilon_{\rm tr}<4t-\Delta$, the polarization changes sign continuously, denoting spin mixing. In the rest of the paper, a constant charge bias $\mu_{\rm L} - \mu_{\rm R} =10^{-2}t$ is applied to the system. We take $L=W$ unless explicitly specified and all lengths are expressed in units of the lattice parameter $a_0$.
For different transport energy of incoming electrons $\epsilon_{\rm tr}$, we plot in Fig. \ref{fig:2}(b) and (c) the topological Hall angle $\theta_{\rm TH}$ and topological spin Hall angle $\theta_{\rm TSH}$ for two different values of exchange coupling $\Delta=\frac{1}{3}t$ and $\frac{4}{3}t$. A global analysis of Fig. \ref{fig:2} shows three main regions, irrespective of the exchange strength:
\begin{enumerate}
\item $|\epsilon_{\rm tr}|>4t-\Delta$, the material is fully spin polarized, $\theta_{\rm TH}$ is negative and finite whereas $\theta_{\rm TSH}$ is zero.
\item $\Delta<|\epsilon_{\rm tr}|<4t-\Delta$, the spin polarization is smaller than 1 and vanishes in most of the region; there $\theta_{\rm TH}=0$ whereas $\theta_{\rm TSH}$ is finite and negative.
\item $|\epsilon_{\rm tr}|<\Delta$, $\theta_{\rm TSH}\simeq 0$ and $\theta_{\rm TH}$ is constant and positive.
\end{enumerate}
The dependence of the topological Hall angles $\theta_{\rm T(S)H}$ on transport energy can be understood by considering the spin and carrier type (electron/hole) injected from the ferromagnetic contacts as explained in Ref. \onlinecite{Yin2015} [see Fig. 2(d) in Ref. \onlinecite{Yin2015}]: for positive bias voltage, electrons are injected from lead L into lead R and holes are injected from lead R into lead L. Under skyrmion-driven topological Hall effect, a spin-up electron originating from lead L scatters towards lead T, and by symmetry a spin-down hole originating from lead R scatters towards lead B. Similarly, a spin-down electron originating from lead L scatters towards lead B, and a spin-up hole originating from lead R scatters towards lead T. We can now analyze the results displayed in Fig. \ref{fig:2}.
In region (1), $|\epsilon_{\rm tr}|>4t-\Delta$, the leads are half metallic so that only spin-up are available. Electrons are scattered towards lead T, while holes are scattered towards lead B and as a result only THE survives while TSHE is quenched ($\theta_{\rm TSH} = 0$). In region (2), both spin-up and spin-down electrons (holes) are injected from terminal L (R). Spin-up and spin-down carriers experience a topological spin-dependent force, $F_\uparrow=-F_\downarrow$, that drags them towards opposite directions. In addition, due to the zero current condition imposed on leads T and B, the diffusion-driven force reacting to charge imbalance is non-topological and spin-independent. Hence, it exerts the {\em same} force on spin-up and spin-down, i.e. $-eE_{\rm TH}=F_\uparrow=F_\downarrow$. As a consequence, these two conditions are met only when $\theta_{\rm TH}=0$. In region (3), spin-down electrons and holes are injected from terminals L and R, respectively, so that two different types of carriers with the same spin dominate the transport. TSHE is suppressed, and THE becomes finite.

As a final note, we stress out that our calculations are performed on large samples and therefore account for a large number of modes. When the calculation is performed in a narrow sample displaying a small number of modes, as in Ref. \onlinecite{Yin2015}, it results in the manifestation of quantum interferences yielding oscillations of the T(S)HE signal as a function of the energy. Such oscillations are unlikely to be observed in a realistic situation due to decoherence. The large number of modes accounted for in our study ensures that the computed T(S)HE signals are smooth, free from quantum oscillations, and hence correspond to a more realistic experimental situation. 

\subsection{The validity of the adiabatic approximation and the geometrical Hall signals}
The theory of THE has been mostly derived within the adiabatic approximation, i.e. assuming that the flowing spins remain aligned on the local magnetization \cite{Bruno2004,Nagaosa2013}. Nonetheless, when the magnetic texture changes abruptly (typically on a distance equivalent to the spin precession length) the itinerant spins start misaligning away from the local magnetization, an effect known as the spin mistracking and responsible for domain wall resistance and non-adiabatic torque \cite{Tatara2004,Xiao2006}. In the present section, we aim at determining whether the result obtained from the adiabatic theory is valid in few-atom-size skyrmions. \par
\begin{figure}[t]
\includegraphics[width=8.4cm]{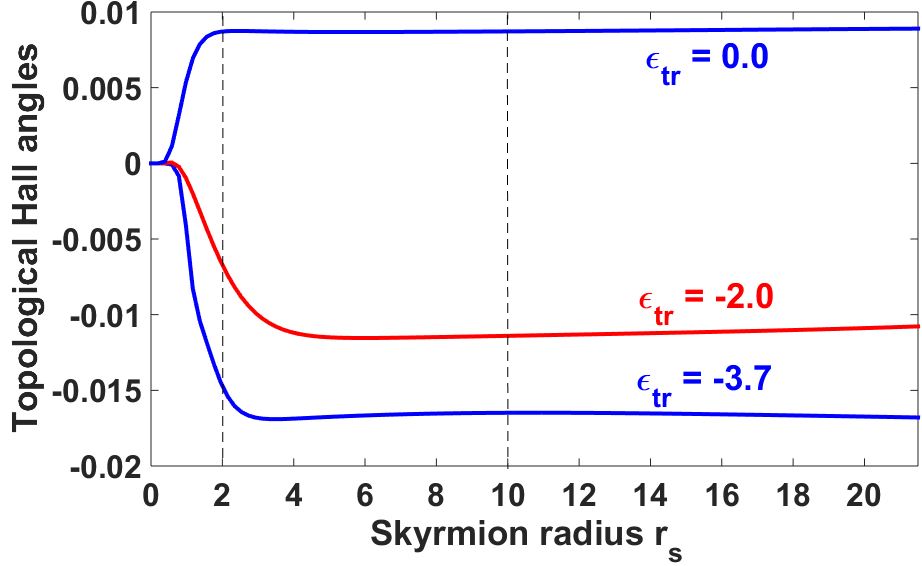}
\caption{\label{fig:3} (Color online) The non zero topological Hall angles $\theta_{\rm TH}$ (blue) and $\theta_{\rm TSH}$ (red) as a function of the skyrmion radius for three different energies, $\epsilon_{\rm tr}=-3.7$, $-2.0$ and 0. Here, the sample size is 128$\times$128 $a_0^2$ and the exchange coupling constant is $\Delta=\frac{2t}{3}$. The vertical lines indicate the minimal and the optimal skyrmion radius.}
\end{figure}
To do so, we compute THE and TSHE as a function of the skyrmion size in a sample of width $W=128$ $a_0$. Notice that the maximal skyrmion radius is taken at r$_s=\frac{W}{6}$ to avoid spurious size effects and unwanted magnetic discontinuity at the edges of the sample. 
The numerical ansatz for the skyrmionic profile is very convenient for this purpose because for this range of r$_s$ the magnetization is almost fully relaxed to an up state at the leads. The results are plotted in Fig. \ref{fig:3} for different transport energies: the blue curves represents the nonvanishing THE for $\epsilon_{\rm tr}=0$, $-3.7$ and the red one the TSHE at $\epsilon_{\rm tr}=-2.0$. The value r$_s$ = 0 corresponds obviously to the absence of skyrmion, i.e. the homogeneous ferromagnetic state, and does not display any THE or TSHE as seen in Fig. \ref{fig:3}. When the radius increases, the system gradually departs from the ferromagnetic state and a single skyrmion is generated so that the topological Hall angles increase from 0 to a finite value. The radius r$_s=a_0$ corresponds to a single spin down impurity in the middle of the ferromagnetic state and therefore does not represent a true skyrmion. But above the critical radius of r$_s=2a_0$, our model captures a proper skyrmion and both THE and TSHE saturate at a constant value, independent of the skyrmion radius.  As a matter of fact, a small skyrmion occupies a narrow region but exhibits a large emergent magnetic field, due to the large magnetization gradient. Hence, although only few electrons experience the emergent magnetic field, they are strongly deflected. On the other hand, a large skyrmion presents a much smaller emergent magnetic field due to its weak magnetization gradient, but occupies a much wider region of space. Therefore, almost all electrons are (weakly) deflected.
This balance between strength of emergent field and number of deflected electrons explains the constant value observed in Fig. \ref{fig:3}. 
That is the reason why the theory expresses THE and TSHE as a function of the magnetic flux and not the magnetic field. We conclude that the adiabatic approximation assumed in the conventional theories of THE \cite{Bruno2004,Nagaosa2013} is very robust, even for very small skyrmions (two atomic sites). For the rest of the paper, we consider a fixed skyrmion radius equal to r$_0$ = 10$a_0$.\par

\begin{figure}[t]
\includegraphics[width=8.4cm]{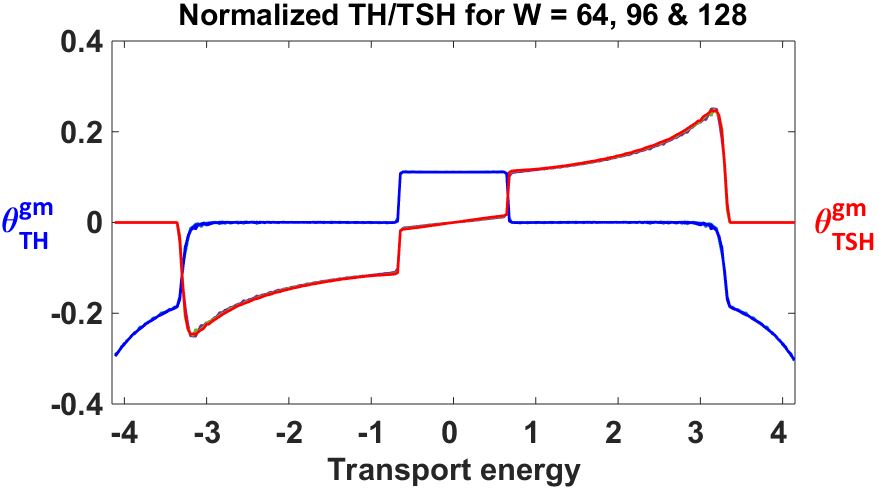}
\caption{\label{fig:4} (Color online) The normalized $\theta_{\rm TH}^{\rm gm}$ (blue) and $\theta_{\rm TSH}^{\rm gm}$ (red) as a function of $\epsilon_{\rm tr}$. The skyrmion radius is r$_s$ = 10$a_0$. For all widths $W=L=$ 64$a_0$, 96$a_0$ and 128$a_0$, the normalized signals are superimposed and give exactly the same value. For reference, the exchange coupling is $\Delta=\frac{2t}{3}$.}
\end{figure} 

Finally, we show that our results do not depend on the system size and are therefore independent of the number of modes. Phenomenological reasoning suggests that the longitudinal conductance increases with the sample size while the transverse (topological Hall) conductance is only given by the skyrmion and remains constant as a function of the width. Hence, by applying the appropriate scaling transformation, the resulting T(S)HE curves should all superpose, irrespective of the width of the sample. 
Fig. \ref{fig:4} displays the geometrical T(S)HE, defined \begin{equation}
\theta_{\rm T(S)H}^{\rm gm} =\frac{W}{2{\rm r}_0}\cdot\theta_{\rm T(S)H}\nonumber
\end{equation} as a function of the transport energy for various sample sizes, $W =$ 64$a_0$, 96$a_0$ and 128$a_0$. As expected, all curves superpose with each other demonstrating that our results are free from spurious quantum interferences and that the sample boundaries have no impact on our numerical results.

\subsection{Robustness of Topological Hall Signals}
 \begin{figure}[b]
\includegraphics[width=8.1cm]{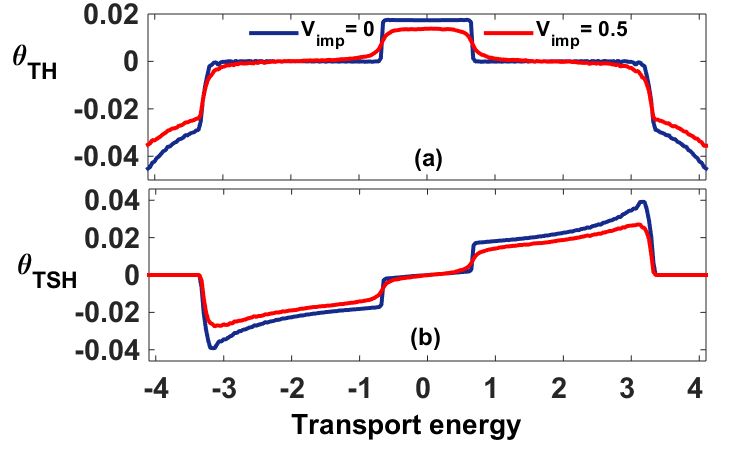}
\caption{\label{fig:5} (Color online) Energy dependence of (a) $\theta_{\rm TH}$ and (b) $\theta_{\rm TSH}$ in the clean limit and for $V_{\rm imp}=0.5$, $\Delta=\frac{2t}{3}$, and $W=L=$ 64$a_0$.}
\end{figure} 

So far in this study, we have assumed ballistic transport in clean regime. It has been recently shown that momentum scattering against defects and impurities has a dramatic impact on spin transport in any realistic magnetic textures \cite{Yuan2012,Yuan2016}. As a matter of fact, since spin transport in magnetic textures presents striking similarities with spin transport in spin-orbit coupled band structure, momentum scattering breaking the coherent spin precession around the local magnetic field results, for instance, in enhanced non-adiabaticity parameter \cite{Akosa2015}. Consequently, one expects that impurity scattering is detrimental to the skyrmion induced Hall effects studied above. The aim of this section is to provide some insight on the robustness of T(S)HE in disordered skyrmionic textures. The impurities are numerically introduced in our two-dimensional square lattice by adding a spin-independent random potential $\mathcal{V}_i^{\rm si}$ to the onsite energy $\epsilon_0$ , such that $\mathcal{V}_i^{\rm si} \in [-\frac{V_{\rm imp}}{2}, \frac{V_{\rm imp}}{2}]$, where $V_{\rm imp}$ defines the disorder strength. Figures \ref{fig:5}(a) and (b) display the TH/TSH angles in presence (red line) and absence of impurities (blue line): it is shown that disorder smears out the edge and boundaries and reduces the magnitude of Hall signals. \par
 
\begin{figure}[t]
\includegraphics[width=8.4cm]{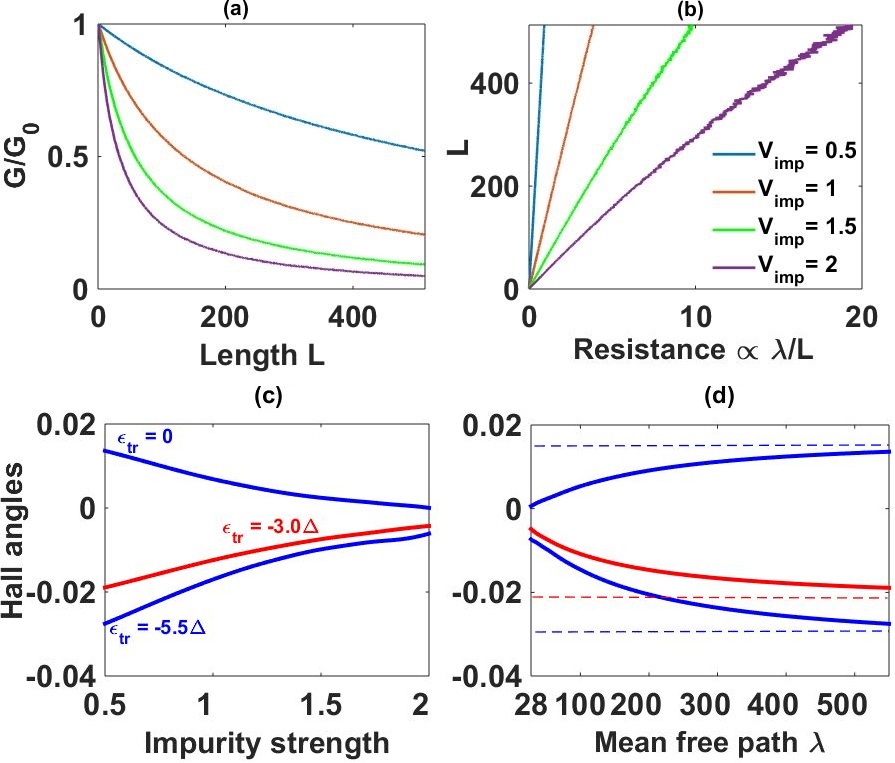}
\caption{\label{fig:6}(Color online) (a) Conductance of the metallic layer as a function of its length for different impurity strengths. (b) Extraction of the mean free path $\lambda$ for the first transport energy $\epsilon_{\rm tr}=-3.7$. Note that changing the transport energy modifies the correspondance between $V_{\rm imp}$ and $\lambda$. (c,d) The non vanishing $\theta_{\rm TH}$ (blue) and $\theta_{\rm TSH}$ (red) for a sample size of $W=64$ and for three energies, (c) as a function of impurity strength $V_{\rm imp}$ and (d) as a function of the equivalent mean free path $\lambda$.}
\end{figure}

For further physical insight, we systematically vary the impurity strength over a wide range $V_{\rm imp}\in[0,2]t$. In order to quantify the impact of disorder on T(S)HE, we first express the impurity strength in terms of its equivalent mean free path $\lambda$. To do so, we calculate the conductance of the two-terminal sample, keeping its width fixed at $W=64a_0$ and varying its length $L$ for different disorder strengths $V_{\rm imp}$. The curves of the normalized conductance are shown in Fig. \ref{fig:6}(a). We then extract the mean free path corresponding to each disorder strength following the semiclassical formula of the conductance,
\begin{equation}
G=G_0 \bigg/ \left(1+\frac{L}{\lambda}\right),
\end{equation}
where $G_0=(e^2/h)N$ with $N$ standing for the number of transport modes in the sample. The resistance of the sample and its length follow approximately a linear relationship, the proportionality constant allowing to extract the effective mean free path of the system, as shown in Fig. \ref{fig:6}(b). For $V_{\rm imp}$ vayring from $0.5t$ to $2t$, the equivalent mean free path varies from 25$a_0$ to 550$a_0$. The localization effects are negligible here. 

Figure \ref{fig:6}(c,d) displays the topological Hall angles as a function of (c) the impurity strength and (d) the equivalent mean free path. The dashed lines in Fig. \ref{fig:6}(d) indicate the values of the T(S)HE in the clean limit. These calculations demonstrate clearly that THE and TSHE are very sensitive to disorder. As a matter of fact, although the skyrmion radius is quite small, ${\rm r}_s=10a_0$ in this calculation, the topological Hall angles are reduced by about 50\% for a mean free path about 20 times the skyrmion radius.

\section{Conclusion and Perspectives}
The topological properties of electronic transport in skyrmionic textures have been investigated in the clean and disordered regimes. In particular, we showed that the relative strength of the topological Hall and topological spin Hall effects can be discriminated according to the energy of incoming electrons and exchange coupling. 
The optimal size of the sample and of the skyrmion maximizes the magnitude of the Hall angles, the scale being determined by the geometrical topological Hall angles. Finally, the robustness of these effects with respect to spin-independent impurity scattering is quite weak, as the topological Hall angles are quenched for a mean free path much larger than the skyrmion size. 
\acknowledgments
The authors acknowledge fruitful discussions with A. Abbout, S. Ghosh and Z. T. Ndiaye. This work was supported by the King Abdullah University of Science and Technology (KAUST) through the Award No OSR-CRG URF/1/1693-01 from the Office of Sponsored Research (OSR).


%

\end{document}